\documentclass[10pt,aps,showpacs,a4paper,floatfix,twocolumn,tightenlines]{revtex4}
\usepackage{epsfig}
\usepackage{psfig}
\usepackage{graphicx}
\usepackage{graphics}
\usepackage{xspace}
\usepackage{amssymb}
\usepackage{amsmath}
\usepackage{latexsym}
\usepackage{natbib}
\usepackage{mathrsfs}
\newcommand{\half}{\textstyle{1\over 2}}
\newcommand{\thalf}{\textstyle{3\over 2}}

\newcommand{\sla}{\mskip 1.mu /\mskip-9mu}      
\newcommand{\ove}{\overline}                    
\newcommand{\inieq}{\begin{eqnarray}}            
\newcommand{\fineq}{\end{eqnarray}}            

\def\p{\mbox{\boldmath $p$}}

\def\q{\mbox{\boldmath $q$}}

\def\ss{\mbox{\boldmath $\sigma$}}
\def\ta{\mbox{\boldmath $\tau$}}

\def\pf{\mbox{\boldmath $p^{ \prime}$}}
\def\pm{\mbox{\boldmath $p_{\mathrm m}$}}

\def\ep{\mbox{$e^{\prime}$}}

\begin{document}
\title{ Meson exchange currents in a relativistic model for electromagnetic one 
nucleon emission }

\author{Andrea Meucci, Carlotta Giusti, and Franco Davide Pacati}
\affiliation{Dipartimento di Fisica Nucleare e Teorica, 
Universit\`{a} di Pavia, and \\
Istituto Nazionale di Fisica Nucleare, 
Sezione di Pavia, I-27100 Pavia, Italy}

\date{\today}

\begin{abstract}
We analyze the role of meson exchange currents (MEC) in photon- and 
electron-induced
one nucleon emission reactions in a fully relativistic model. The 
relativistic mean field theory is used for the bound state
and the Pauli reduction for the scattering state. Direct 1-body and exchange
2-body terms in the nuclear current are considered. Results for the 
$^{12}$C$\left(\gamma,p\right)$ and $^{16}$O$\left(\gamma,p\right)$ 
differential cross sections and photon
asymmetries are displayed in an energy range between 60 and 196 MeV. The 2-body 
seagull current affects the cross section less than in nonrelativistic 
analyses. In the case of the
$^{16}$O$\left(\gamma,n\right)$ differential cross section MEC effects 
are large but not sufficient to reproduce the data. MEC have a small 
effect on $\left(e,e^{\prime}p\right)$ calculations.
\end{abstract}

\pacs{ 25.20.Lj, 25.30.Fj, 24.10.Jv, 24.10.Eq}

\maketitle


\section{introduction}
\label{intro}

One nucleon knockout reactions are a primary tool to explore the 
single-particle
aspects of the nucleus. Several measurements at different energies and
kinematics have been performed in a wide range of target nuclei, which 
stimulated the
production of a considerable amount of theoretical calculations. 

The validity of the direct knockout (DKO) mechanism is clearly
established for exclusive $\left(e,e^{\prime}p\right)$ reactions~\cite{book}. 
Theoretical models based on the nonrelativistic and relativistic distorted wave 
impulse 
approximation (DWIA) are able to give an excellent description of data in a 
wide range of nuclei and in different kinematics. In contrast, the reaction 
mechanism of photonuclear reactions has been the object of a longstanding 
discussion~\cite{book}. On the one hand, the DKO mechanism, with a suitable 
choice of the theoretical ingredients adopted for bound and scattering states, 
was able to describe $\left(\gamma,p\right)$ cross sections for photon energies 
up to $E_{\gamma} \simeq$ 100 MeV~\cite{Boffi81}. On the other hand, the fact 
that the 
transitions with neutron emission are of the same order of magnitude as those 
with proton emission addressed to a reaction mechanism where the transferred 
momentum is shared between two nucleons. Indeed the quasi-deuteron 
model~\cite{Levinger,Schoch,Eden} was applied with some success to 
photoreactions at low and medium energy. Various corrections were included in 
the DKO model~\cite{Boffi84,Boffi85}, but were unable to give a consistent
description of $\left(\gamma,p\right)$ and $\left(\gamma,n\right)$ data.

In recent years, tagged photon facilities were developed and produced data with
high energy resolution and a clear separation between different states of the
residual nucleus~\cite{spring,miller,Abeele,Andersson,deBever,Branford}. For the
$\left(\gamma,p\right)$ reaction, various analyses in different theoretical
approaches suggest that the DKO contribution may be a small fraction of
data~\cite{Ireland,bobe,mori}, thus indicating that a prominent role is played
by more complicated mechanisms, such as meson exchange currents (MEC) and
multistep processes due to nuclear correlations. 

Nonrelativistic DWIA calculations with
ingredients for bound and scattering states consistent with
$\left(e,e^{\prime}p\right)$ reactions are unable to describe
$\left(\gamma,p\right)$ data~\cite{Benenti,Gaid}. A reasonable agreement is
obtained when the MEC contribution is added to the DKO. MEC are found to produce
an enhancement of the DKO cross sections~\cite{Benenti}.
The importance of MEC in proton photoemission was also studied in 
Ref.~\cite{ryck3} for the $^{12}$C$\left(\gamma,p\right)$ reaction at 
intermediate energy. 

Isobar current (IC) effects in photonuclear reactions were studied in
Ref.~\cite{bright1}, where a microscopic calculation including both nuclear
correlations and $\Delta$ excitations showed that IC are small except at large
momentum transfer. The model was then extended to include also MEC and applied
to proton capture $\left(p,\gamma\right)$ in Ref.~\cite{bright2} and
suggested that the DKO is the most important contribution to this reaction. The
role of MEC and $\Delta$ excitations in $\left(\gamma,p\right)$ reactions was
analyzed in Ref.~\cite{co}, where also short range correlations were considered.
Large differences between DKO cross sections and those obtained with the
inclusion of MEC were found for large proton emission angles. 

The relativistic approach was first applied to $\left(\gamma,p\right)$ reactions
in Ref.~\cite{mcder}, where also MEC were considered, and in 
Refs.~\cite{lotz,Joha} within the framework of DKO. The DKO mechanism was able 
to reproduce the $^{16}$O$\left(\gamma,p\right)$ data at 
$E_{\gamma} = 60$ MeV~\cite{Joha}. The same approach was then extended in 
Ref.~\cite{joha1} to a much wider energy range and showed that the DKO is the 
main contribution to the cross section for missing momentum values up to 
$p_m \simeq$ 500 MeV/$c$, while MEC and IC are expected to give important 
effects for larger missing momenta. 

The effects of MEC and IC in $\left(e,e^{\prime}p\right)$ reactions at
quasielastic peak were first presented within a nonrelativistic framework in 
Ref.~\cite{br}, where a small contribution of MEC and a reduction due to IC were
obtained. In contrast, in Ref.~\cite{ryck1}, important effects on the
interference response functions were found out. Moreover, the effects were
dependent on the shell considered. The sensitivity of polarization observables
to MEC and IC in $\left(\vec{e},e^{\prime}\vec{p}\right)$ was studied in
Ref.~\cite{ryck2}, where a moderate dependence on MEC was predicted only at
$p_m \gtrsim$ 200 MeV/$c$. In Ref.~\cite{amaro}, MEC 
and IC effects on $\left(e,e^{\prime}p\right)$ are generally small.

Different fully relativistic DWIA (RDWIA) models were developed in recent years
and successfully applied to the analysis of $\left(e,e^{\prime}p\right)$
data~\cite{RDWIA,RDWIA2,meucci1}. In a recent paper~\cite{meucci2}, we have 
compared
nonrelativistic and relativistic calculations for the $\left(\gamma,N\right)$
knockout reactions in order to clarify the relationship between the DWIA and
RDWIA approaches for $\left(\gamma,p\right)$ and $\left(\gamma,n\right)$, and
to study the relevance of the DKO mechanism in nonrelativistic and relativistic
calculations. In this work our interest is focused on the role played by MEC
in $\left(\gamma,N\right)$ and in $\left(e,e^{\prime}p\right)$ reactions
within the framework of RDWIA. 

The RDWIA treatment is the same as in Ref.~\cite{meucci2}. The relativistic 
bound state wave functions are solutions of a Dirac equation 
containing scalar and vector potentials obtained in the framework of the 
relativistic mean field theory.
The effective Pauli reduction has been adopted for the outgoing nucleon wave 
function. This simple scheme is in principle equivalent to the 
exact solution of
the Dirac equation. The resulting Schr\"odinger-like equation is
solved for each partial wave starting from relativistic optical potentials.
The same spectroscopic factors obtained in Refs.~\cite{meucci1,rmd} by fitting 
our RDWIA $\left(e,e^{\prime}p\right)$ results to data have 
been applied to the calculated $\left(\gamma,N\right)$ cross sections. 

Results for $^{12}$C and $^{16}$O target nuclei at different photon energies
have been considered. The 1-body part of the relativistic current is written 
following the most commonly used current conserving (cc) prescriptions 
for the $\left(e,e^{\prime}p\right)$ reaction introduced in Ref.~\cite{deF}. 
The ambiguities connected with different choices of the electromagnetic current 
cannot be dismissed. In the $\left(e,e^{\prime}p\right)$ reaction the 
predictions of different 
prescriptions are generally in close agreement~\cite{pollock}. Large differences
can however be found at high missing momenta~\cite{off1,off2}. 
These differences are increased in $\left(\gamma,N\right)$ reactions,
where the kinematics is deeply off-shell and higher values of the missing 
momentum are probed. 

The 2-body part of the current is constructed starting from
the pseudovector $\pi N$ Lagrangian as in Refs.~\cite{donn,tjon}. As a first 
step, in this paper we include in the 2-body current only the term
corresponding to the seagull (contact) diagram with one-pion 
exchange. Thus, we consider
only a part of the contribution of MEC. This contribution, however, should be 
able to understand the relevance of the 2-body currents in a relativistic
approach also in comparison with previous nonrelativistic 
calculations.  

The formalism is outlined in Sec.~\ref{sec.for}. Relativistic calculations of 
the $^{12}$C$\left(\gamma,p\right)$ and $^{16}$O$\left(\gamma,p\right)$ cross 
sections are presented in Sec.~\ref{sec.results}, where also MEC effects on the 
$(\gamma,n)$ and $\left(e,e^{\prime}p\right)$ reactions are discussed. Some 
conclusions are drawn in Sec.~\ref{con}.   

\section{formalism}
\label{sec.for}

The matrix elements of the nuclear current operator, i.e.,
\inieq
J^{\mu} = \langle \Psi_{\textrm {f}}\mid j^{\mu}\mid \Psi_{\textrm {i}}\rangle 
\ ,\label{eq.jmu}
\fineq
represent the main
ingredient of the cross section and contain all the physical information which
can be extracted from the reaction.

The nuclear current operator can be expanded into 1-body, 2-body, and higher
order components. In this paper 1-body, $j^{\mu}(1\textrm{b})$, and 2-body,  
$j^{\mu}(2\textrm{b})$, terms are included.
The nuclear initial state, $|\Psi_{\textrm {i}}\rangle$, is the many-body
independent-particle model wave function, i.e., a Slater determinant, where only
correlations due to the Pauli principle are included. For exclusive processes
where only one nucleon is emitted and under the assumption that only the
observed channel contributes to the scattering wave function, we can assume that
only one nucleon undergoes a transition and that the residual nucleus is a pure
one-hole state in the target. Then, the matrix elements in Eq.~(\ref{eq.jmu})
are given by the sum of two terms, for the 1-body and the 2-body current
operators, as
\inieq
\langle \Psi_{\textrm {f}}\mid j^{\mu}\mid \Psi_{\textrm {i}}\rangle
\simeq 
\langle \chi^{(-)}(1)\mid j^{\mu}(1\textrm{b})\mid \Psi_{\beta}(1)\rangle 
\nonumber \\ +
\sum_{\alpha = 1}^{\textrm{A}} \langle \chi^{(-)}(1)\Psi_{\alpha}(2)\mid
 j^{\mu}(2\textrm{b})
\mid \Psi_{\beta}(1) \Psi_{\alpha}(2)
\nonumber \\ 
- \Psi_{\alpha}(1) \Psi_{\beta}(2) \rangle
\ , \label{eq.mat}
\fineq  
where $\chi^{(-)}$ is the distorted wave function of the emitted nucleon, 
and $\Psi_{\alpha(\beta)}$ are single-particle bound state wave functions. 

In the first term the interaction occurs, through a 1-body current, only 
with the nucleon that is ejected and the other nucleons behave as spectators. 
This term corresponds to the DKO mechanism  and gives the RDWIA. In the second 
term the interaction occurs, through a 2-body current, with a pair of 
nucleons. Only one nucleon is emitted and the other nucleon of the pair is 
reabsorbed in the residual nucleus. For the nucleon which is not emitted a 
sum over all the the single-particle states is performed in the calculations.  

At present, there is no unambiguous approach for dealing with off-shell
nucleons. Here, we discuss the three cc expressions for the 1-body current
\cite{deF,kelly2,meu}
\begin{eqnarray}
 j_{\textrm{cc}1}^{\mu} &=& G_M(Q^2) \gamma ^{\mu} - 
             \frac {\kappa}{2M} F_2(Q^2)\overline P^{\mu} \ , \nonumber \\
 j_{\textrm{cc}2}^{\mu} &=& F_1(Q^2) \gamma ^{\mu} + 
             i\frac {\kappa}{2M} F_2(Q^2)\sigma^{\mu\nu}q_{\nu} \ ,
	     \label{eq.cc} \\
 j_{\textrm{cc}3}^{\mu} &=& F_1(Q^2) \frac{\overline P^{\mu}}{2M} + 
             \frac {i}{2M} G_M(Q^2)\sigma^{\mu\nu}q_{\nu} \ , \nonumber
\end{eqnarray}
where $q^{\mu} = (\omega,\q)$ is the four-momentum transfer,
$Q^2=|\q|^2-\omega ^2$, $\overline P^{\mu} = (E+E',\pm+\pf)$, $E'$ and $\pf$ are 
the energy and momentum of the emitted nucleon, $\kappa$ is the anomalous part 
of the magnetic
moment, $F_1$ and $F_2$ are the Dirac and Pauli nucleon form factors, $G_M =
F_1+\kappa F_2$ is the Sachs nucleon magnetic form factor, and
$\sigma^{\mu\nu}=\left(i/2\right)\left[\gamma^{\mu},\gamma^{\nu}\right]$. These 
expressions are equivalent for on-shell particles due to Gordon identity,
but they give different results when applied to off-shell nucleons. 

The 2-body current is due to meson exchanges between nucleons. We have 
considered in this paper only the seagull 
diagram. The corresponding current is written in momentum space 
as \cite{donn,tjon}
\inieq
J_{\textrm{S}}^{\mu} = - F_{\textrm{S}} \frac{f^2}{m_{\pi}^2}
\ove{\Psi}(1)\gamma^{\mu}\gamma^5\Psi(1) \ove{\Psi}(2)\sla k_2 \gamma^5 \Psi(2)
\nonumber \\ 
\frac{1}{k_2^2-m_{\pi}^2} \xi_1^{\dagger}\xi_2^{\dagger} i (\ta_1 \times
\ta_2)_z \xi_1\xi_2 + (1\leftrightarrow 2) \ , \label{eq.seag}
\fineq
where $F_{\textrm{S}} = G_E^p - G_E^n$, $f^2/(4\pi) \simeq 0.079$, 
$m_{\pi} \simeq 140$ MeV is the pion mass, and $\xi$ is the isospin wave 
function. We have performed calculations with the cutoff $\Lambda = 1250$
MeV in the pion propagator.

Current conservation is restored by replacing the longitudinal current and the 
bound nucleon energy by  \cite{deF}
\begin{eqnarray}
J^L &=& J^z = \frac{\omega}{\mid\q\mid}~J^0 \ , \\
E &=& \sqrt{\mid \pm \mid^2 + M^2} = \sqrt{ \mid \pf-\q\mid^2 + M^2} \ .
\end{eqnarray}
The bound state wave functions 
\inieq
\Psi_{\alpha(\beta)} = \left(\begin{array}{c} u_{\alpha(\beta)}  \\ 
 v_{\alpha(\beta)} \end{array}\right) \ , \label{eq.bwf}
\fineq
are given by the Dirac-Hartree solution of a relativistic Lagrangian
containing scalar and vector potentials. 

The ejectile wave function  is written in terms of its positive energy
component following the direct Pauli reduction scheme~\cite{HPa}, i.e.,
\inieq
\chi = \left(\begin{array}{c} \chi_{+} \\ \frac {\ss\cdot\p'}{M+E'+S-V}
        \chi_{+} \end{array}\right) \ ,
\fineq
where $S=S(r)$ and $V=V(r)$ are the scalar and vector potentials for the nucleon
with energy $E'$. The upper component, $\chi_{+}$, is related to a
Schr\"odinger equivalent wave function, $\Phi_{f}$, by the Darwin 
factor, $D(r)$, i.e.,
\inieq
\chi_{+} &=& \sqrt{D(r)}\Phi_{f} \ , \\
D(r) &=& 1 + \frac{S-V}{M+E'} \ .
\fineq
$\Phi_{f}$ is a two-component wave function which is solution of a 
Schr\"odinger
equation containing equivalent central and spin-orbit potentials obtained from
the scalar and vector potentials.

The coincidence cross section of the $\left(e,\ep p\right)$ reaction can be 
written in terms of four response functions,
$f_{\lambda\lambda^{\prime}}$, as
\inieq
\sigma  &=& \sigma _{\mathrm M}f_{\mathrm {rec}} E' |\pf |\ \left\{\rho _{00}
f_{00} +  \rho _{11}f_{11}+\rho _{01}f_{01}\cos\left(\vartheta\right)
\right. \nonumber
\\ &+&
 \left. \rho _{1-1}f_{1-1}\cos\left(2\vartheta\right)\right\} \ ,  
\label{eq.fcs}
\fineq
where $\sigma _{\mathrm M}$ is the Mott cross section, $f_{\mathrm {rec}}$ is 
the recoil factor \cite{book,kellyrep}, and $\vartheta $ is the out-of-plane 
angle between the electron scattering plane and the $(\q, \pf)$ plane. The 
coefficients $\rho_{\lambda\lambda'}$ are obtained from the lepton tensor
components and depend only upon the electron kinematics \cite{book,kellyrep}.

In case of an incident photon with energy $E_{\gamma}$, the 
$\left(\gamma,N\right)$ cross section can be written in terms of the pure
transverse response, i.e.,
\inieq
\sigma_{\gamma} = \frac{2\pi^2\alpha}{E_{\gamma}}\ f_{\mathrm {rec}}\ 
E' |\pf |\  f_{11} \ ,  \label{eq.gcs}
\fineq
where $\alpha \simeq 1/137$. If the photon beam is linearly polarized, also the
interference transverse-transverse response is non-zero and appears in the
definition of the photon asymmetry
\inieq 
A = -\frac{f_{1-1}}{f_{11}} \ . \label{eq.asy}
\fineq
The response functions are given by bilinear combinations of the nuclear current
components, i.e.,
\inieq
f_{00} &=& \langle J^0 \left(J^0\right)^{\dagger }\rangle \ , \nonumber \\ 
f_{11} &=& \langle J^x \left(J^x\right)^{\dagger} \rangle +
          \langle J^y \left(J^y\right)^{\dagger} \rangle \ , \nonumber \\
f_{01} &=& -2\sqrt 2\ \mathfrak{Re}
   \left[\langle J^x \left(J^0\right)^{\dagger }\rangle\right] \ , \nonumber \\
f_{1-1} &=&  \langle J^y \left(J^y\right)^{\dagger} \rangle -
          \langle J^x \left(J^x\right)^{\dagger} \rangle \ ,
\fineq
where $\langle\cdots\rangle$ means that average over the initial and sum over
the final states is performed fulfilling energy conservation. 

\section{results and discussion}
\label{sec.results}

The results of this section have been obtained with the same
bound state wave functions and optical potentials as in
Refs.~\cite{meucci1,meucci2}, where the RDWIA 1-body analysis was successfully 
applied to reproduce $\left(e,e^{\prime}p\right)$ and $\left(\gamma,p\right)$ 
data. 

The relativistic bound state wave functions have been obtained from the code of
Ref.~\cite{adfx}, where relativistic Hartree-Bogoliubov equations are solved in
the context of a relativistic mean-field theory that satisfactorily reproduces
single-particle properties of several spherical and deformed nuclei~\cite{lala}. 
The direct Pauli reduction is applied for the scattering state which is 
calculated by means of
the energy- and mass number-dependent complex phenomenological optical potential
EDAD1 of Ref.~\cite{chc}. The EDAD1 potential is obtained from fits to proton
elastic scattering data on several nuclei in an energy range up to 1040 MeV.
Since there is no unambiguous prescription for handling off-shell nucleons, we 
have performed calculations with different cc expressions for the 1-body 
current. The Dirac and Pauli form factors are taken from Ref.~\cite{mmd}.

\subsection{The \bf{$\left(\gamma,p\right)$} and \bf{$\left(\gamma,n\right)$}
reactions}

The analysis of $\left(\gamma,p\right)$ reactions has been the object of a
longstanding discussion about the reaction mechanism. Many nonrelativistic
calculations in different theoretical approaches suggested that MEC and
$\Delta$ excitations should play a prominent role. On the contrary, the
RDWIA approach seems to indicate that the DKO
mechanism is the leading process, at least for low photon energies and missing
momenta up to $\simeq 500$ MeV$/c$. Our aim is to study whether
this conclusion is correct investigating the effects of the seagull (SEAG) 
current on the cross section.
  

\begin{figure}
\includegraphics[height=10cm, width=8.4cm]{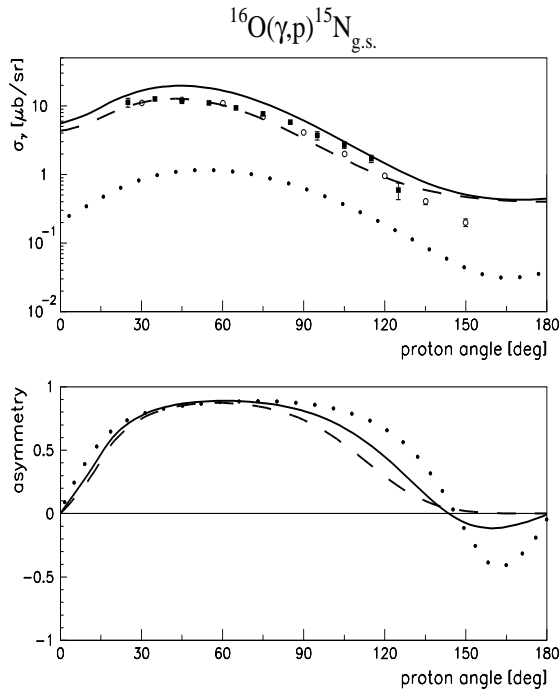} 
\caption {The cross section and photon asymmetry for the
$^{16}$O$\left(\gamma,p\right)^{15}$N$_{\textrm{g.s.}}$ reaction as functions 
of the proton scattering angle at $E_{\gamma} = 60$ MeV. The data are from
Ref.~\cite{miller} (black squares) and from Ref.~\cite{findlay} (open circles).
Solid lines represent the DKO+SEAG results, dashed lines the DKO results, and
dotted lines the SEAG results. }
\label{fig:fig1}
\end{figure}



\begin{figure}
\includegraphics[height=10cm, width=8.4cm]{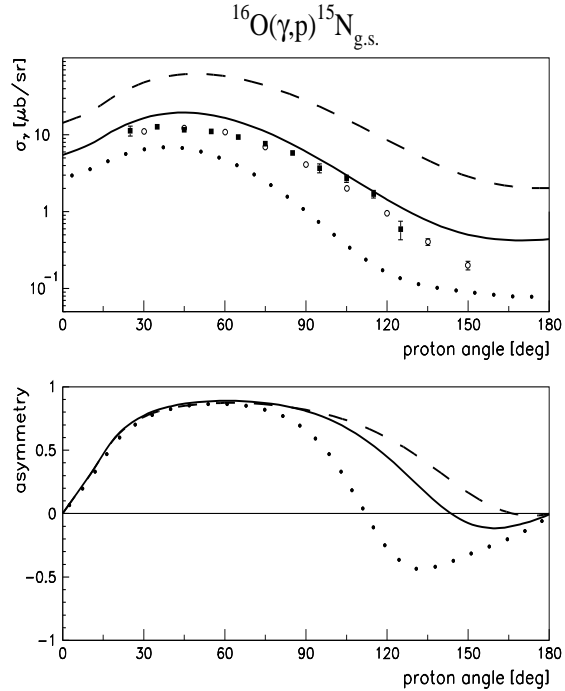} 
\caption {The cross section and photon asymmetry for the
$^{16}$O$\left(\gamma,p\right)^{15}$N$_{\textrm{g.s.}}$ reaction as functions 
of the proton scattering angle at $E_{\gamma} = 60$ MeV. The data are from
Ref.~\cite{miller} (black squares) and from Ref.~\cite{findlay} (open circles).
Dashed, solid, and dotted lines represent the DKO+SEAG results, with cc1, cc2,
and cc3 prescriptions for the 1-body current, respectively. }
\label{fig:fig2}
\end{figure}


The comparison between the DKO+SEAG, DKO, and SEAG results is shown in
Fig.~\ref{fig:fig1} for the cross section and photon asymmetry of the
$^{16}$O$\left(\gamma,p\right)^{15}$N$_{\textrm{g.s.}}$ reaction at
$E_{\gamma} = 60$ MeV. The cc2 current has been used and the spectroscopic
factor $Z(p\half)=0.71$ has been applied \cite{meucci1,meucci2,rmd}. 
As it was already known
from previous analyses~\cite{joha1,meucci2}, the 1-body term provides the main 
contribution to the cross section and can satisfactorily reproduce the
data, at least for small angles. The pure contribution of the 2-body term 
is one order of magnitude lower than the 1-body one, but their interference 
is large. The total result is enhanced above the
data and the shape is slightly affected. The SEAG contribution 
is sizable but less than in previous nonrelativistic
calculations~\cite{Benenti}. It has been pointed out in a nonrelativistic
approach~\cite{co} that the SEAG term overestimates 
MEC. A substantial reduction 
is obtained when the pion-in-flight diagram is added, while the $\Delta$ 
current is important only with increasing photon energies. If these results 
were confirmed in relativistic calculations, the pion-in-flight term would
reduce the contribution of seagull and bring the calculated cross 
section in Fig.~\ref{fig:fig1} closer to the DKO results and also to the data. 
 
The photon asymmetry at $E_{\gamma} = 60$ MeV is
shown in the lower panel of Fig.~\ref{fig:fig1}. The differences between the
DKO+SEAG and the DKO results are generally small, but at large angles, 
where the SEAG contribution becomes negative.

The sensitivity of the $\left(\gamma,p\right)$ calculations at 
$E_{\gamma}=60$ MeV to different cc prescription for the 1-body current is 
presented in Fig.~\ref{fig:fig2}, where results for the DKO+SEAG contribution 
are displayed.
As we already pointed out in Ref.~\cite{meucci2}, large differences are 
given by the three expressions of the 1-body current at the considered
photon energy. These differences are somewhat reduced when the seagull current
is added, but remain anyhow large. The calculated cross sections are strongly 
enhanced if we use cc1; this is probably due to an overestimation of the 
convective current contribution for an off-shell nucleon. Results with cc3 are 
lower than those with cc2, but the difference decreases with increasing photon
energy. Large differences are obtained also on the photon asymmetry at large 
scattering angles. 


\begin{figure}
\includegraphics[height=10cm, width=8.4cm]{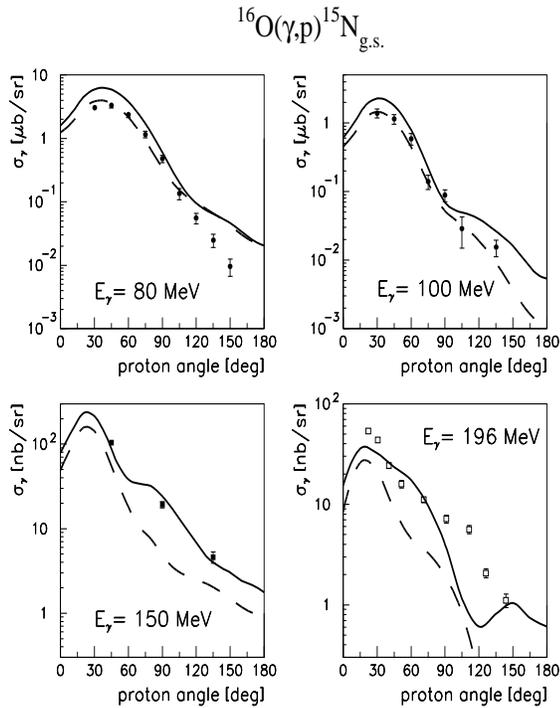} 
\caption {The cross section for the
$^{16}$O$\left(\gamma,p\right)^{15}$N$_{\textrm{g.s.}}$ reaction as a function 
of the proton scattering angle at photon energy ranging from 80 to 196 MeV. The 
data at 80 and 100 MeV are from Ref.~\cite{findlay}. The data at 150 MeV are 
from Ref.~\cite{leitch}. The data at 196 MeV are from Ref.~\cite{adams}. 
Solid lines represent the DKO+SEAG results and dashed lines the DKO results.}
\label{fig:fig3}
\end{figure}



\begin{figure}
\includegraphics[height=10cm, width=8.4cm]{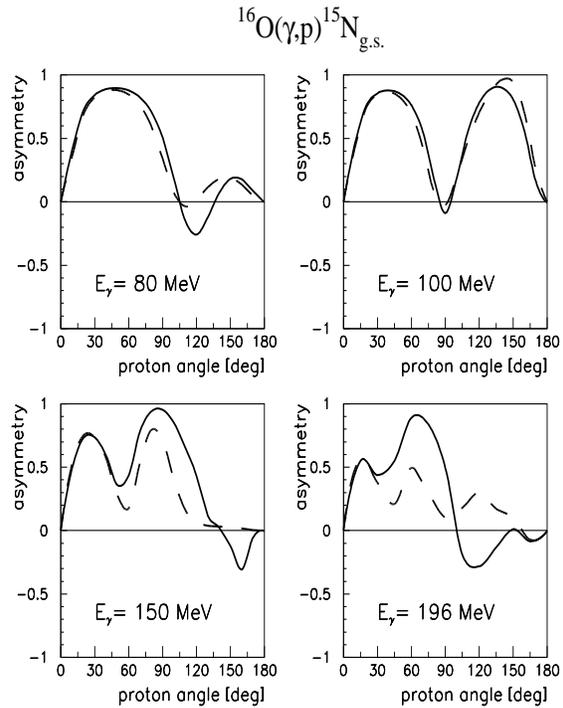} 
\caption {The same as in Fig.~\ref{fig:fig3}, but for the photon asymmetry.}
\label{fig:fig4}
\end{figure}


In Figs.~\ref{fig:fig3} and~\ref{fig:fig4} the comparison between the 
DKO+SEAG and DKO results is shown for the cross section and the photon 
asymmetry for energy ranging from 80 to 196 MeV. The seagull contribution 
enhances the cross section at all the considered photon energies. Thus, the
experimental cross sections at $E_{\gamma}=80$ and 100 MeV, that are already 
reproduced by the DKO result, are overestimated, while a better agreement with 
data is found at $E_{\gamma}=150$ and 196 MeV. In order to draw definite 
conclusions in comparison with data, however, it would be useful to check 
the relevance of the pion-in-flight contribution and also of the 
IC, that should play a significant role above 150 MeV.
For the photon asymmetry in Fig.~\ref{fig:fig4} the differences between the 
DKO+SEAG and DKO results increase with the scattering angle and with the 
photon energy.


\begin{figure}
\includegraphics[height=10cm, width=8.4cm]{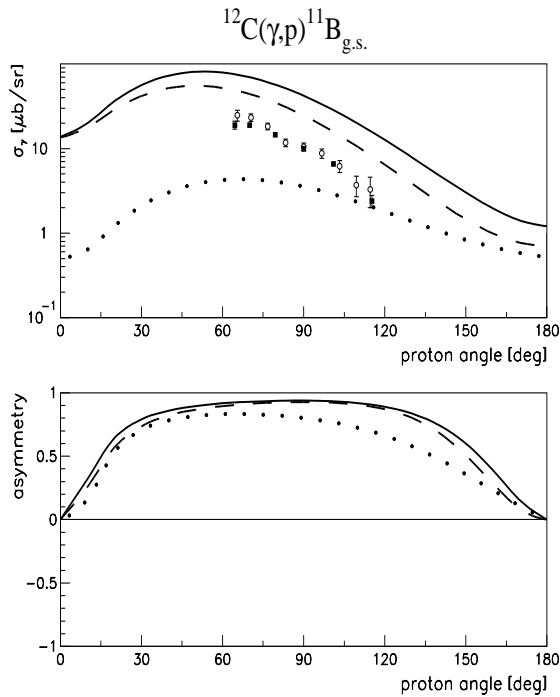} 
\caption {The cross section and photon asymmetry for the
$^{12}$C$\left(\gamma,p\right)^{11}$B$_{\textrm{g.s.}}$ reaction as functions 
of the proton scattering angle at $E_{\gamma} = 58.4$ MeV. The data are from
Ref.~\cite{shotter} (black squares) and from Ref.~\cite{spring} (open circles).
Line convention as in Fig.~\ref{fig:fig1}.}
\label{fig:fig5}
\end{figure}


In Fig.~\ref{fig:fig5} the cross section and the photon asymmetry for the 
$^{12}$C$\left(\gamma,p\right)^{11}$B$_{\textrm{g.s.}}$ reaction at 
$E_{\gamma}=58.4$ MeV are presented. The spectroscopic factor $Z(p\thalf)=0.56$ 
has been applied. Also in this case, the DKO+SEAG results are greater 
than the
DKO ones. However the most apparent feature is that none of them can reproduce
the data. This fact was already found out in Refs.~\cite{joha1,meucci2}, where
it was suggested that a better agreement might be obtained with a more clear
determination of the $^{12}$C ground state, which should take into account its
intrinsic deformation.


\begin{figure}
\includegraphics[height=10cm, width=8.4cm]{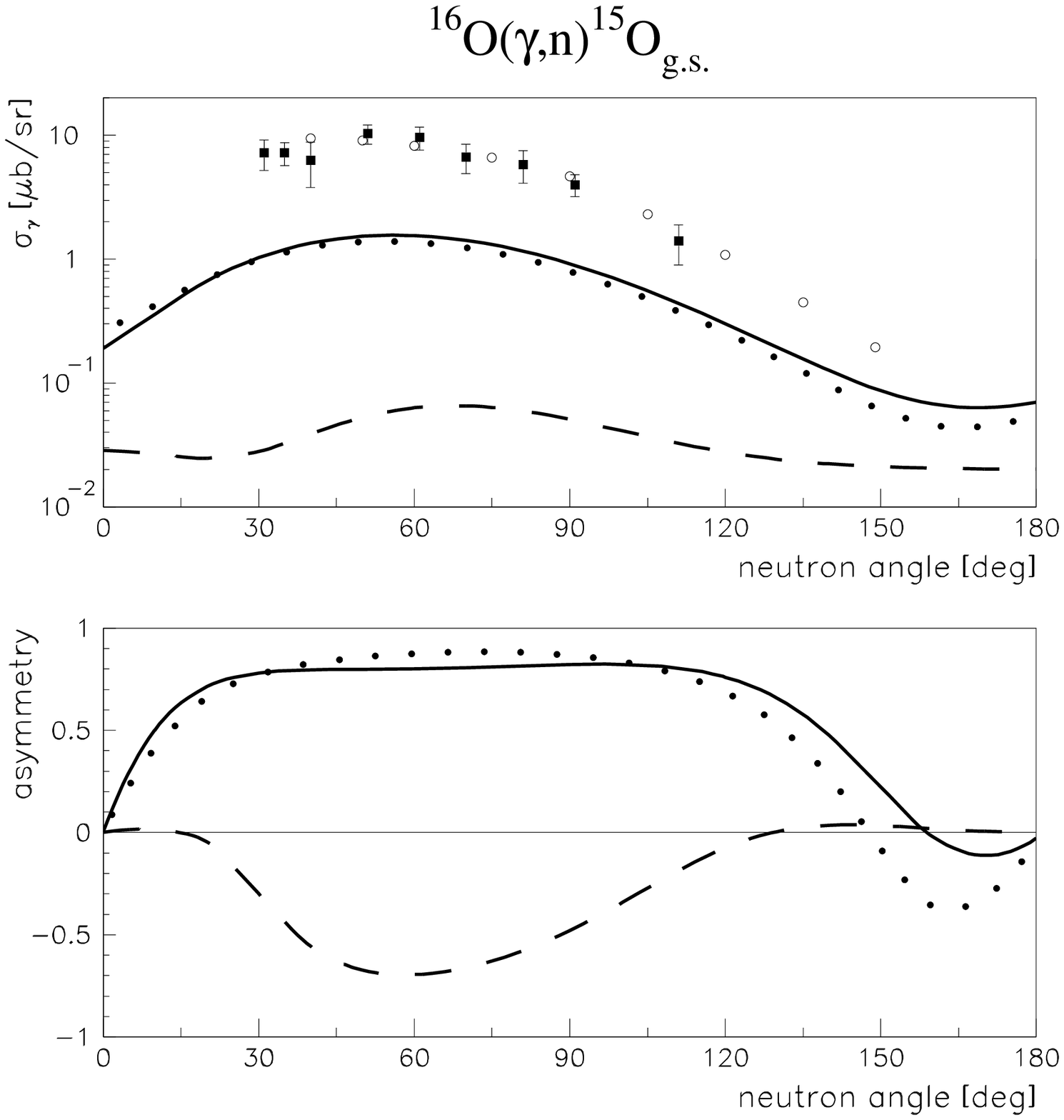} 
\caption {The cross section and photon asymmetry for the
$^{16}$O$\left(\gamma,n\right)^{15}$O$_{\textrm{g.s.}}$ reaction as functions 
of the neutron scattering angle at $E_{\gamma} = 60$ MeV. The data are from
Ref.~\cite{Andersson} (black squares) and from Ref.~\cite{goringer} (open 
circles). Line convention as in Fig.~\ref{fig:fig1}. }
\label{fig:fig6}
\end{figure}


Results for neutron photoemission
at $E_{\gamma}=60$ MeV are displayed in Fig.~\ref{fig:fig6}. The same
spectroscopic factor as in the $\left(\gamma,p\right)$ reaction has been 
applied.
The fact that the ratio between experimental $\left(\gamma,p\right)$ and
$\left(\gamma,n\right)$ cross sections is comparable to unity has been
traditionally interpreted as a signal of the dominance of a 2-body mechanism in
the $\left(\gamma,n\right)$ reaction. We see that results with DKO+SEAG are
greatly increased with respect to the DKO ones, but this enhancement is still 
insufficient to reproduce the magnitude of the data. These
results seem to indicate that more complicated effects are needed to reproduce
the data, such as, e.g., a rescattering 
process \cite{bright1,ryck4,Andersson,co}.


\begin{figure}[th]
\includegraphics[height=10cm, width=8.4cm]{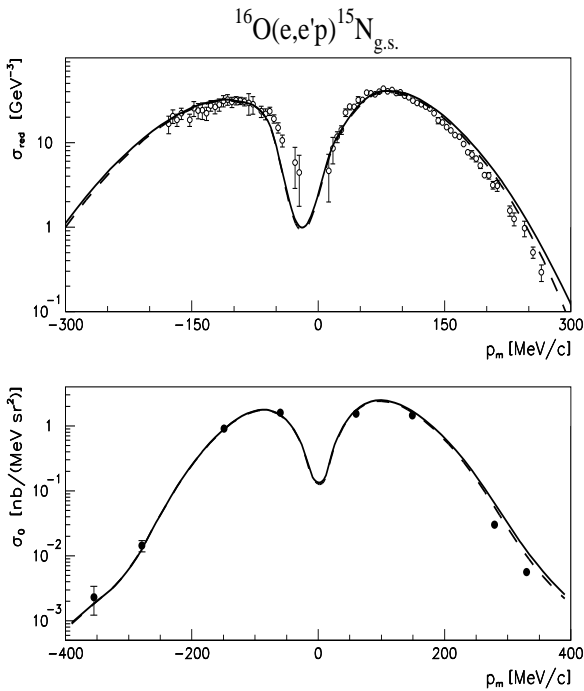} 
\caption {Upper panel: reduced cross section for the
$^{16}$O$(e,e'p)^{15}$N$_{\textrm{g.s.}}$
reaction at $E_p = 90$ MeV constant proton energy in the center-of-mass 
system in parallel kinematics~\cite{nikhef}. Lower panel: 
cross section for the same reaction, but at $Q^2=0.8$ (GeV/$c)^2$ in 
constant $\left(\q,\omega\right)$ kinematics~\cite{e89003}. Solid lines 
represent the DKO+SEAG results and dashed lines the DKO results.}
\label{fig:fig7}
\end{figure}


\subsection{The {\bf$\left(e,e^{\prime}p\right)$} reaction}

The study of the exclusive $\left(e,e^{\prime}p\right)$ knockout 
reaction for
$Q^2\leq 0.4$ (GeV/$c)^2$ was successfully performed in the theoretical
framework of nonrelativistic DWIA. In more recent years, owing to the new data
at $Q^2\simeq 0.8$ (GeV/$c)^2$ from Jefferson Laboratory
(JLab)~\cite{e89003,malov}, models based on a fully relativistic approach were
developed. Both nonrelativistic and relativistic $\left(e,e^{\prime}p\right)$ 
analyses were performed including the 1-body current only. In fact, the 2-body 
diagrams were not expected to give an important contribution, at least over the
explored kinematics conditions. 

In Fig.~\ref{fig:fig7} the 
$^{16}$O$\left(e,e^{\prime}p\right)^{15}$N$_{\textrm{g.s.}}$ reaction is
considered. In the upper panel the reduced cross section data measured at
NIKHEF~\cite{nikhef} in parallel kinematics with a proton energy of 90 MeV in 
the center-of-mass system are compared with our DKO+SEAG and DKO
calculations. The cc2 prescription for the 1-body current has been used and the
spectroscopic factor is $Z(p\half)=0.71$. In the lower panel the same reaction
is studied at the JLab constant $\left(\q,\omega\right)$ 
kinematics~\cite{e89003}.
As it was already found in Ref. \cite{meucci1}, the DKO calculation gives
good descriptions of the data in both kinematics. A slight enhancement is
due to the seagull current and is visible only at higher values of $p_m$. This 
result is consistent with usual expectations for which quasifree
electron scattering is almost unaffected by MEC. 

We have also performed calculations for the transition to the 
$p\thalf$ first excited state of $^{15}$N at the
same kinematics as in Fig.~\ref{fig:fig7} but we have not found any appreciable
difference with respect to the $p\half$ state. We have
also calculated the response functions measured in 
$^{16}$O$\left(e,e^{\prime}p\right)^{15}$N at JLab~\cite{e89003}
and the polarization observables from MIT-Bates~\cite{batespn} 
on $^{12}$C$\left(e,e^{\prime}\vec{p}\right)^{11}$B 
and JLab~\cite{malov} on $^{16}$O$\left(\vec{e},e^{\prime}\vec{p}\right)^{15}$N.
MEC might be expected to give a more significant effect in the induced
polarization, but we have not found any significant difference with
respect to our RDWIA results of Refs. \cite{meucci1,rmd}.

\section{Summary and conclusions}
\label{con}

In this paper a first step has been made to study the role of MEC in 
$\left(\gamma,N\right)$ and $\left(e,e^{\prime}p\right)$ reactions in a 
fully relativistic framework. In previous relativistic and nonrelativistic DWIA
calculations the DKO mechanism was clearly established for quasifree 
$\left(e,e^{\prime}p\right)$ reactions in comparison with data, and only a small
contribution is expected from 2-body currents. Various nonrelativistic 
calculations give different results, but confirm that the contribution of
MEC in $\left(e,e^{\prime}p\right)$ is not very important. 
Nonrelativistic analyses of $\left(\gamma,p\right)$ reactions generally indicate
a prominent role of MEC. Their contribution is important to reproduce the data 
and  affects the shape and size of the calculated cross sections at all the 
photon energies. 
In contrast, RDWIA calculations suggest that the DKO mechanism is already able
to give a reasonable agreement with data and MEC seem to be required only at
$p_m \gtrsim$ 500 MeV/$c$. Thus, our aim was to study the relevance of
2-body currents in comparison with DKO within a fully relativistic framework.   

The nuclear current operator is expanded into 1-body and 2-body components.
The 1-body term gives the DKO contribution. For the 2-body term we assume 
that only a pair of nucleons are involved in the reaction: one is emitted from 
a specific state and the other one is reabsorbed in the nucleus, i.e., the 
residual nucleus is a one-hole state in the target. 

In the transition matrix elements of the nuclear current operator the bound 
state wave function is obtained in the framework of the relativistic mean 
field theory, and the direct Pauli reduction method with scalar and vector 
potentials is used for the ejectile wave functions. In order to study the 
ambiguities in the 1-body electromagnetic vertex due to the off-shellness 
of the initial nucleon, we have performed calculations using three current 
conserving expressions. 

As a first step, we have considered in this paper only 
the contribution to the MEC due to the seagull diagram.
We have discussed the effect of this term 
on the $\left(\gamma,p\right)$ reactions for photon energies up to 196
MeV. As in previous RDWIA analyses, the DKO term
provides the main contribution to the cross section and is in satisfactory 
agreement with the data, at least for small energies and angles. The pure SEAG 
term is smaller than the DKO one. The total effect enhances the cross 
section, but less than in nonrelativistic calculations. 
On the other hand, in nonrelativistic calculations the 
pion-in-flight diagram reduces the effect of the seagull 
current, while the $\Delta$ excitation is
important only with increasing photon energies. 
In the case of our RDWIA calculation, we expect a similar result. The  
inclusion of all MEC contributions should have a more limited but still visible 
effect on the cross section, while IC
should become important at increasing photon energies. 

Large ambiguities to the different prescriptions for the 1-body current are
generally found in the $\left(\gamma,p\right)$ cross section also when the 
seagull current is included.

For the $\left(\gamma,n\right)$ reaction, 
the dominant contribution of a 2-body mechanism has been traditionally claimed
to explain the magnitude of the experimental cross section. Our RDWIA results
are greatly increased when the SEAG contribution is included, but the 
enhancement is still insufficient to reproduce the data. This seems to 
indicate that more complicated effects are needed to reproduce
the data. A careful and consistent analysis of these mechanisms in a 
relativistic framework would be important and helpful to clarify this 
question. 

We have also performed calculations for the 
$\left(e,e^{\prime}p\right)$ reaction at different kinematics.
Also in this case, the seagull diagram enhances the RDWIA results,
but, in contrast to $\left(\gamma,p\right)$, the effects 
are generally small and visible only at high missing momenta. Thus, the
comparison with data, that were already well reproduced by the DKO model, is
practically unaffected.

\end{document}